\documentclass[12pt]{article}

\usepackage{graphicx}
\usepackage{dcolumn}
\usepackage{bm}
\usepackage{soul}
\usepackage{ulem}
\usepackage{color}
\usepackage[T1]{fontenc}
\usepackage[super]{nth}
\usepackage{amsmath}
\usepackage[center]{subfigure}
\usepackage[utf8]{inputenc}
\usepackage{array}
\usepackage{tabu}
\usepackage[super]{nth}
\usepackage{tikz}
\usepackage{standalone}
\usepackage{amssymb}
\usepackage{cite}
\usepackage{float}

\def\beq{\begin{equation}}
\def\eeq{\end{equation}}
\def\ba{\begin{eqnarray}}
\def\ea{\end{eqnarray}}

\begin{document}

\begin{center}
{\Large{\bf Finite-Range Coulomb Gas Models II: Applications to Quantum Kicked Rotors and Banded Random Matrices}} \\
\ \\
\ \\
by \\
Avanish Kumar, Akhilesh Pandey and Sanjay Puri \\
School of Physical Sciences, Jawaharlal Nehru University, New Delhi -- 110067, India.
\end{center}

\begin{abstract}
In paper I of this two-stage exposition, we introduced finite-range Coulomb gas (FRCG) models, and developed an integral-equation framework for their study. We obtained exact analytical results for $d=0,1,2$, where $d$ denotes the range of eigenvalue interaction. We found that the integral-equation framework was not analytically tractable for higher values of $d$. In this paper II, we develop a Monte Carlo (MC) technique to study FRCG models. Our MC simulations provide a solution of FRCG models for arbitrary $d$. We show that, as $d$ increases, there is a transition from Poisson to Wigner-Dyson classical random matrix statistics. Thus FRCG models provide a novel route for transition from Poisson to Wigner-Dyson statistics. The analytical formulation obtained in paper I, and MC techniques developed in this paper II, are used to study banded random matrices (BRM) and quantum kicked rotors (QKR). We demonstrate that, for a BRM of bandwidth $b$ and a QKR of chaos parameter $\alpha$, the appropriate FRCG model has range $d=b^2/N=\alpha^2/N$, for $N \rightarrow \infty$. Here, $N$ is the dimensionality of the matrix in BRM, and the evolution operator matrix in QKR. 
\end{abstract}

\newpage

\section{Introduction}
\label{s1}

Random matrices \cite{mm04,bff81,cp65,ew67} have found extensive applications in quantum chaos, i.e., the study of quantum systems whose classical counterpart are chaotic \cite{bg85,bgs84,cc95,fh01,br05,hs06}. The connection between quantum chaos and random matrices is well established. An important paradigm of quantum chaotic systems is the quantum kicked rotor (QKR)
\cite{cc95,fh01,hs06}. The Hamiltonian of the QKR is periodic in time with a delta-function perturbation. In paper I, we introduced and analytically studied finite-range Coulomb gas (FRCG) models which define novel classes of random-matrix ensembles. These are parametrized by the range of eigenvalue interactions, denoted as $d$.

In this paper II, we demonstrate the applicability of FRCG models to quantum chaotic systems. We show that spectral fluctuations of the time evolution operator over one period (i.e., the Floquet operator) of the QKR can be modeled by FRCG models. In other applications, FRCG models have also been used to study quantum pseudo-integrable systems \cite{bgs99,bgs01}. We expect that they would also be applicable to many other physical systems.

An unusual property of quantum chaos is the suppression of chaotic diffusion. In classically chaotic systems such as the classical kicked rotor, the average energy of the system grows linearly with time. However, in its quantum chaotic counterpart (i.e., the QKR), the average energy saturates in time. The suppression of diffusion in the QKR is also known as {\it dynamical localization} \cite{cci79}. The origin of this phenomenon lies in the localization of wave-functions of the Floquet operator in the momentum basis \cite{pf89,fgp82,sm16,gmw98}. In this context, many studies have focused on the transition from ergodic (Wigner-Dyson statistics) to integrable behavior (Poisson statistics) in disordered systems \cite{sm16,gmw98,cb97,am00}. The operator corresponding to the Hamiltonian of quantum systems exhibiting localization can be described by banded random matrices (BRM). Therefore, the eigenvalue statistics of BRM ensembles can also be modeled by FRCG ensembles.

The organization of this paper is as follows. In Sec.~\ref{s2}, we discuss the Monte Carlo (MC) technique for FRCG models. In Sec.~\ref{s3}, we present MC results for FRCG, and compare with analytical results given in Secs.~4-6 of paper I. In Sec.~\ref{s4}, we define QKR and BRM and discuss their connection with FRCG models. We introduce an effective range $d$, which is determined by the parameters of the QKR. In Sec.~\ref{s5}, we present a detailed comparison of FRCG results (analytical and MC) and numerical results for the QKR and BRM. In particular, we will focus on the crossover from Poison to classical ensemble statistics as $d$ increases. In Sec.~\ref{s6}, we discuss FRCG models for fractional values of $d$, and present their applications in QKR. In Sec.~\ref{s7}, we summarize the results presented in this paper.

\section{Monte Carlo Technique for FRCG Models}
\label{s2}

As mentioned in paper I, it becomes harder to obtain complete analytical results for FRCG models with higher values of $d$. However, MC results are easily calculable for arbitrary $d$, thereby providing us a complete picture of FRCG models. In this section, we discuss the MC method. The MC results will supplement our earlier exact results in paper I.

From paper I, we recall the equilibrium joint probability density (jpd) for the linear case:
\begin{equation}\label{e1}
p(x_{1},\cdots,x_{N})= C \exp(-\beta W),
\end{equation}
where the eigenvalues $x_{j}$ are in ascending order. The potential $W$ has a two-body logarithmic potential, and one-body confining potential
\begin{equation}\label{e2}
 W= -{\sum}^{\prime}\log| x_j-x_k|  + \sum_{\substack{j}} V(x_j) .
\end{equation}
Here, ${\sum}^{\prime}$ denotes a sum over all $| j-k| \leq d$ with $j\neq k$, and $d$ denotes the range of the interaction in terms of the particle indices.

The corresponding equilibrium jpd for the circular case is
\begin{equation}\label{e3}
p(\theta_{1},\cdots,\theta_{N})=C \exp\big(-\beta W\big),
\end{equation}
where the eigenangles $\theta_{j}$ are arranged on a unit circle in ascending order. The potential $W$ is given by
\begin{equation}\label{e4}
W= -\frac{1}{2}{\sum}^{\prime}\log| e^{i\theta_{j}}-e^{i\theta_{k}}| + \sum_{\substack{j}}V(\theta_{j}).
\end{equation}
In this case, $V(\theta)$ is a potential periodic on the unit circle. Both the linear and the circular jpds yield the well-known classical ensembles for $d=N-1$. We will shortly describe an MC method for sampling the above jpds.

An equivalent formulation uses the Langevin equation \cite{hr84}. In the linear case, the Langevin  equation for eigenvalues $\lbrace x_{j} \rbrace$ is \cite{ap95}
\begin{equation}\label{e5}
\frac{dx_{j}}{d\tau}= \beta E_{j} + w_{j}(\tau).
\end{equation}
Here, $E_{j}$ is the force arising from the potential in Eq.~(\ref{e2}): 
\begin{equation}\label{e6}
E_{j}= -\frac{\partial W}{\partial x_{j}}.
\end{equation}
The $w_j$ in Eq.~(\ref{e5}) denotes a Gaussian white noise which obeys the appropriate fluctuation-dissipation relation. A similar Langevin equation is obtained for the circular case where $x_{j}\rightarrow \theta_{j}$. One can numerically solve these stochastic differential equations (with adequate precautions to avoid level crossing) to obtain the equilibrium spectra. The Langevin method is not as efficient as the MC method for obtaining equilibrium jpds. However, if one is interested in nonequilibrium evolution also, the Langevin approach will be the method of choice. We will not focus on the Langevin approach in this paper.

\begin{figure}[H]
\centering
\includegraphics[width=0.7\textwidth]{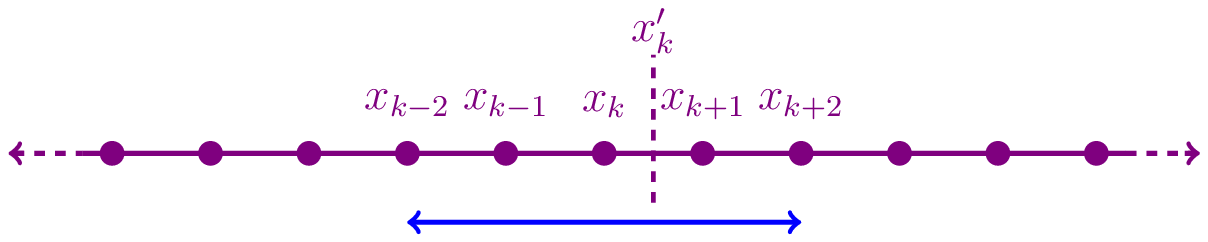}
\caption{Schematic of the MC technique to generate eigenvalue spectra for FRCG models. We consider the case of linear ensembles. The particles on the line denote the positions of the eigenvalues $\{x_j\}$. The eigenvalues interact up to a range $d=2$ in the case shown. This range is measured in terms of the particle indices. In an attempted MC move, the particle $x_k$ is displaced with uniform probability to $x_{k}^{\prime}$ in the range $(x_{k-1},x_{k+1})$. The move is accepted with a probability $\exp(-\beta \triangle W)$, where $\Delta W$ is the change in the potential in Eq.~(\ref{e2}). An MCS corresponds to $N$ attempted moves. The independent spectra are realized by sampling the MC evolution at suitable intervals.}
\label{f1}
\end{figure}

Our MC approach follows \cite{gpp03} for the linear case of Eq.~(\ref{e1}), and \cite{ppk05} for the circular case of Eq.~(\ref{e3}). Because of the logarithmic singularity in the corresponding potentials, particle positions can not change their order. Our MC implementation directly respects this constraint, resulting in a more efficient calculation of the equilibrium jpd (see Fig.~\ref{f1}).

In the \textit{linear case}, we take a set of $N$ eigenvalues  $\left(x_{1},\cdots,x_{N}\right)$ ordered sequentially on a real line with fixed boundaries. The boundaries are chosen such that the probability of finding an eigenvalue outside the range is negligible. A stochastic move assigns, to any randomly chosen $x_{k}$, the new position $x^{\prime}_{k}$ between $(x_{k-1},x_{k+1})$ with a uniform probability. The move is accepted with a probability $\exp(-\beta \triangle W)$, where $\Delta W$ is the change in the potential in Eq.~(\ref{e2}) after the stochastic move. A Monte Carlo step (MCS) corresponds to $N$ attempted moves.

In the \textit{circular case}, we take a set of $N$ eigenvalues $\left(e^{i\theta_{1}},e^{i\theta_{2}},\cdots,e^{i\theta_{N}}\right)$ ordered sequentially on the unit circle. A stochastic move assigns to a randomly chosen eigenangle $\theta_{j}$ the new position $\theta^{\prime}_{j} \in (\theta_{j-1},\theta_{j+1})$ with a uniform probability. After each eigenangle  movement, we use periodic boundary conditions: $\theta^{\prime}_{j}$ is computed modulo $2\pi$. (This respects the original order of their positions on a circle.) The move is accepted with a probability $\exp(-\beta \triangle W)$, where $\Delta W$ is the change in the potential in Eq.~(\ref{e4}) after the stochastic move. 

In Sec.~\ref{s3}, we compare MC and analytical results (from paper I) for the level density and fluctuation measures. We will demonstrate that MC results are in excellent agreement with the analytical results, whenever these are available. Our purpose is to establish the MC technique as a method of obtaining ``exact results'' for higher values of $d$, where analytical results are not available.

All MC results for spectral properties presented in this paper are obtained as averages over $1000$ independent spectra of dimension $N=1001$. For fluctuation measures, we will show results for all three $\beta$-values and the Poisson case ($\beta =0$).

\section{Monte Carlo Results for Level Density and Fluctuation Measures}
\label{s3}

In this section, we present MC results for the level density in linear ensembles. In the uniform circular case, the level density is constant. We also present MC results for the fluctuation measures for many $d$-values, which confirm the transition from Poisson ($d=0$) to classical ($d=N-1$) ensembles.

\begin{figure}[H]
\centering
\includegraphics[width=0.6\textwidth]{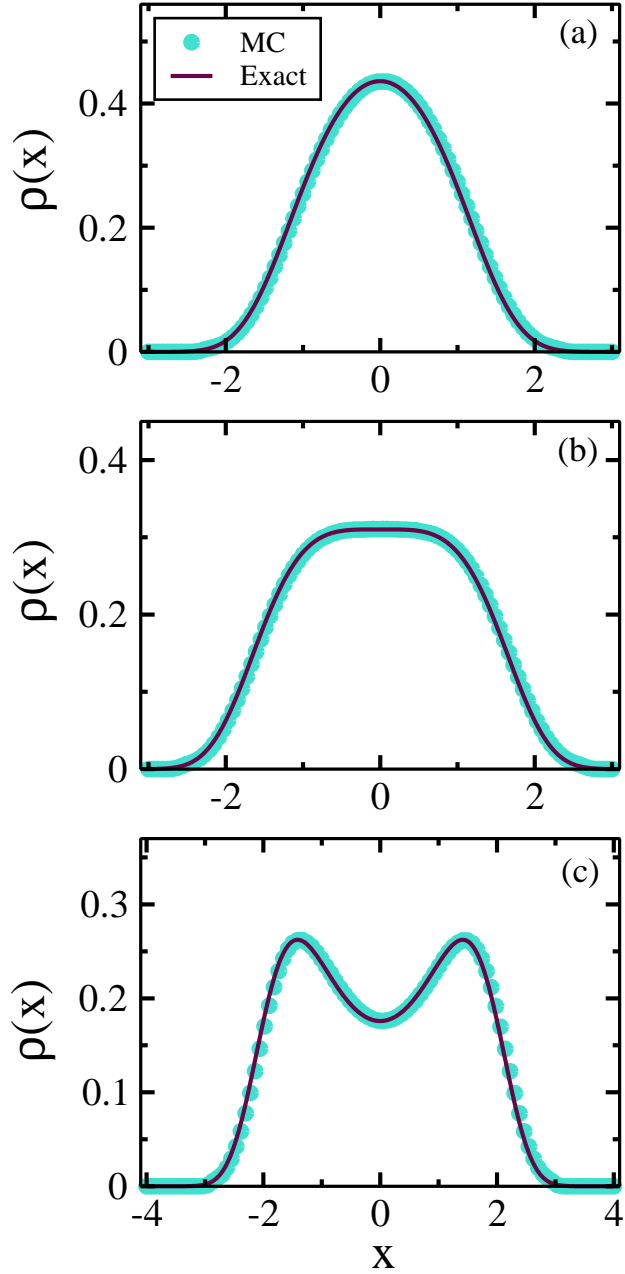}
\caption{Level density for the $d=2$ FRCG model with $\beta=2$. We used the quartic potential with $\kappa=1$, and (a) $\alpha= -2$, (b) $\alpha = 0$, (c) $\alpha = 2$. The filled circles correspond to MC results, and the solid lines are analytical results from Eq.~(29) in paper I.}
\label{f2}
\end{figure}

\begin{figure}[H]
\centering
\includegraphics[width=0.6\textwidth]{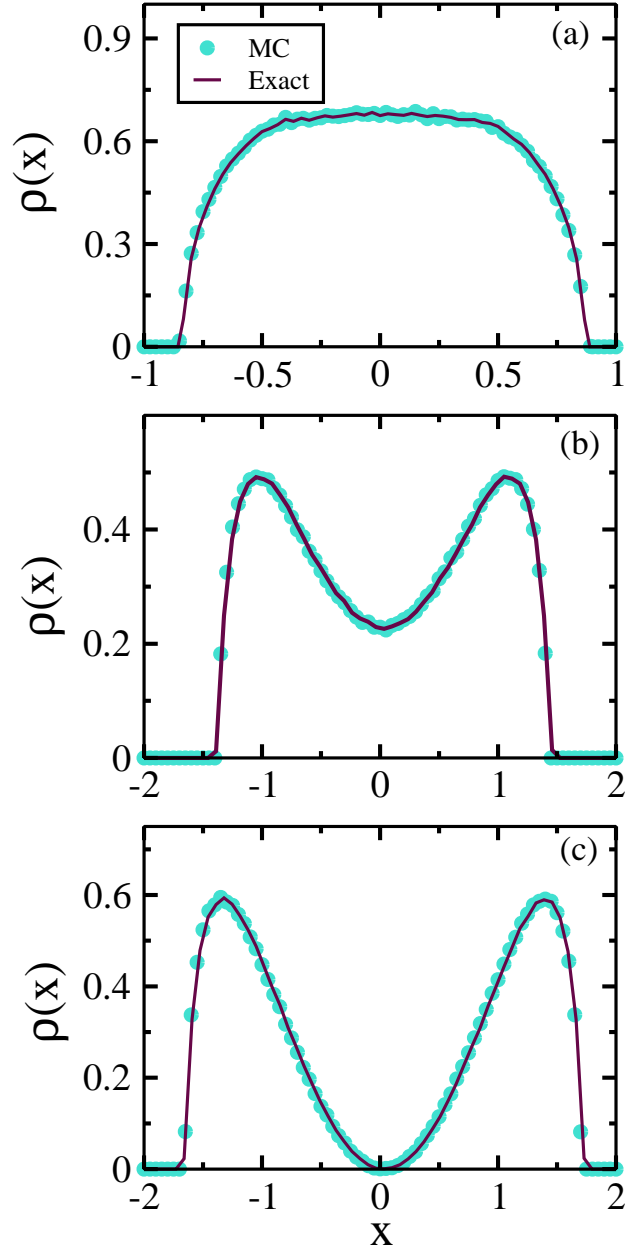}
\caption{Level density for the $d=750$ FRCG model with $\beta=2$. We used the quartic potential with $\kappa = 1$, and (a) $\alpha= -2$, (b) $\alpha = 0$, (c) $\alpha = \alpha_c = 1.414$. The filled circles correspond to MC results, and the solid lines are analytical results from Eqs.~(46)-(47) in paper I.}
\label{f3}
\end{figure}

In paper I, we have presented analytical results for the level density in the linear ensembles with an arbitrary potential. Here, we compare the MC results for the level density with the corresponding exact results for the quartic potential. We have studied this potential for $d = O(1)$ and $d = O(N)$. In each case, excellent agreement is found. Let us show some representative results. We consider the linear case with the quartic potential:
\begin{equation}\label{e7}
V(x) = \kappa\left( \frac{x^{4}}{4}-\alpha\frac{x^{2}}{2}\right), \quad \kappa > 0. 
\end{equation}
Here, $\kappa$ sets the scale of the $V$-axis. In the quartic potential, $\alpha$ determines whether the potential is single well $(\alpha<0)$ or double well $(\alpha>0)$. For $d = N -1$, there is a critical value $\alpha_c$ such that the level density makes a transition from a one-band density $(\alpha <\alpha_{c})$ to a two-band density $(\alpha> \alpha_{c})$ \cite{gpp03}. For $d = O(1)$, the density is given in Eq.~(29) of paper I. This is shown in Fig.~\ref{f2} for $d = 2$. For $d = O(N)$, the density is given in Eqs.~(46)-(47) of paper I (see Fig.~\ref{f3}). Figs.~\ref{f2} and \ref{f3} demonstrate that the level density obtained by the MC technique is numerically indistinguishable from the corresponding analytical results. We have confirmed (not shown here) that the same applies for fluctuation measures for FRCG models with $d=1,2$. Therefore, we will subsequently equate MC results with exact results for FRCG models with arbitrary $d$.

\begin{figure}[H]
\centering
\includegraphics[width=0.6\textwidth]{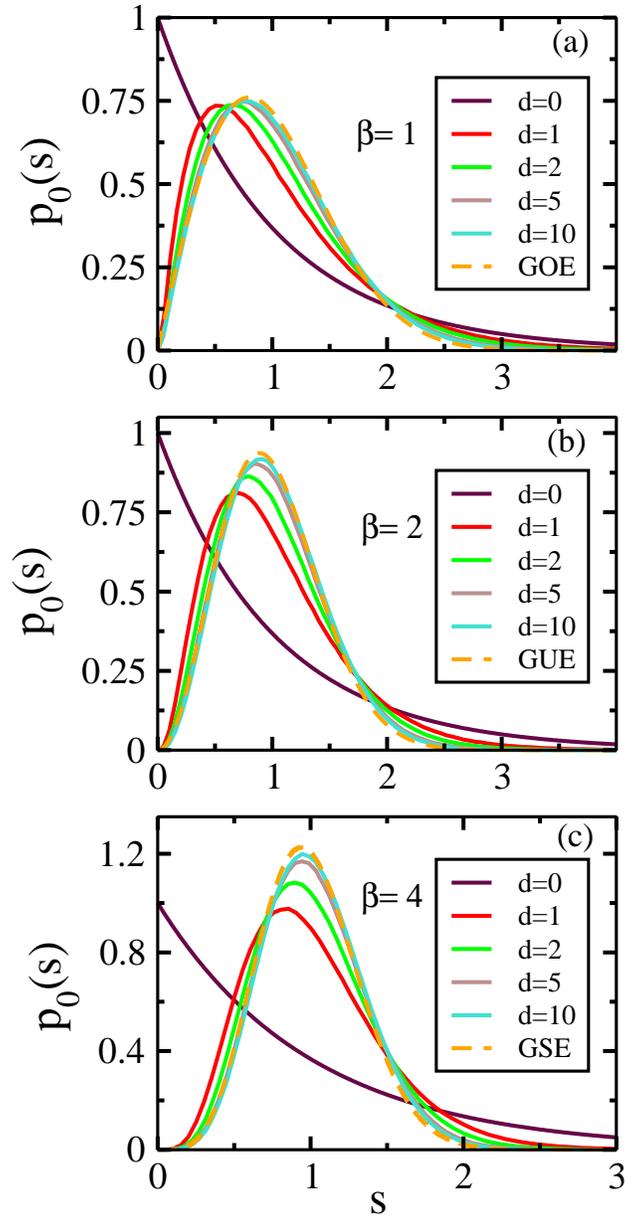}
\caption{Comparison of nearest-neighbor spacing distribution [$p_0(s)$ vs. $s$] for FRCG models with $d=0,1,2,5,10$. We show results for (a) $\beta=1$, (b) $\beta = 2$, and (c) $\beta = 4$. In each frame, the dashed line denotes the classical Gaussian ensemble result.}
\label{f4}
\end{figure}

\begin{figure}[H]
\centering
\includegraphics[width=0.6\textwidth]{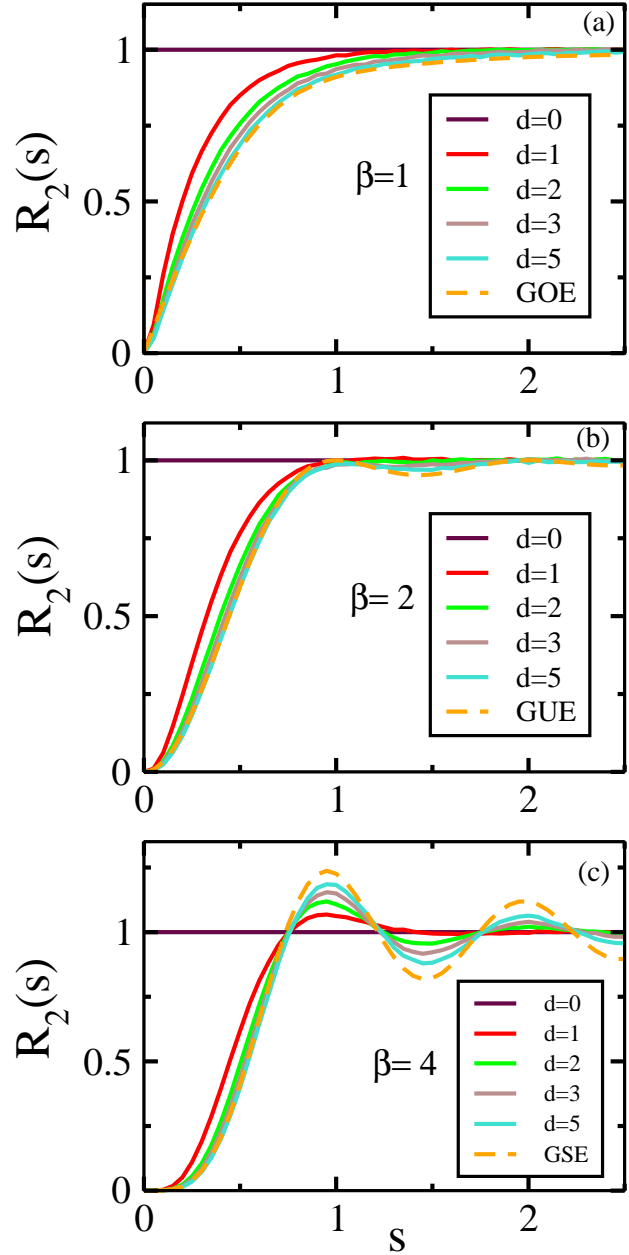}
\caption{Comparison of two-point correlation function [$R_2(s)$ vs. $s$] for FRCG models with $d=0,1,2,3,5$. We show results for (a) $\beta=1$, (b) $\beta = 2$, and (c) $\beta = 4$. In each frame, the dashed line denotes the classical Gaussian ensemble result.}
\label{f5}
\end{figure}

\begin{figure}[H]
\centering
\includegraphics[width=0.6\textwidth]{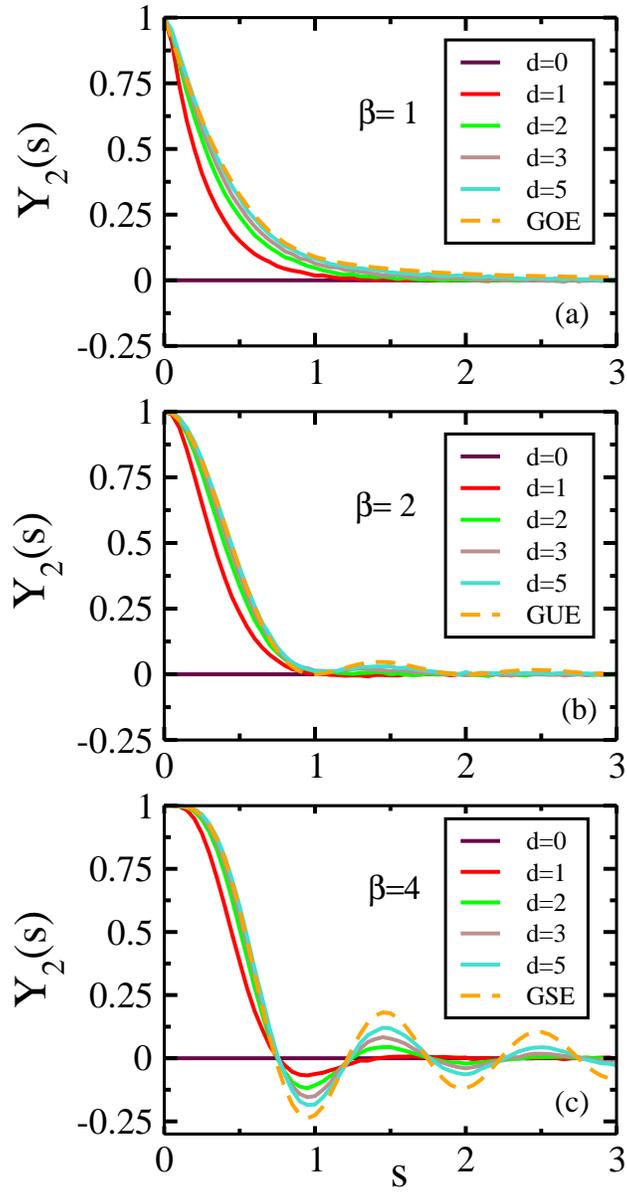}
\caption{Analogous to Fig.~\ref{f5}, but for the two-point cluster function: $Y_2(s) = 1-R_2(s)$.}
\label{f6}
\end{figure}

\begin{figure}[H]
\centering
\includegraphics[width=0.6\textwidth]{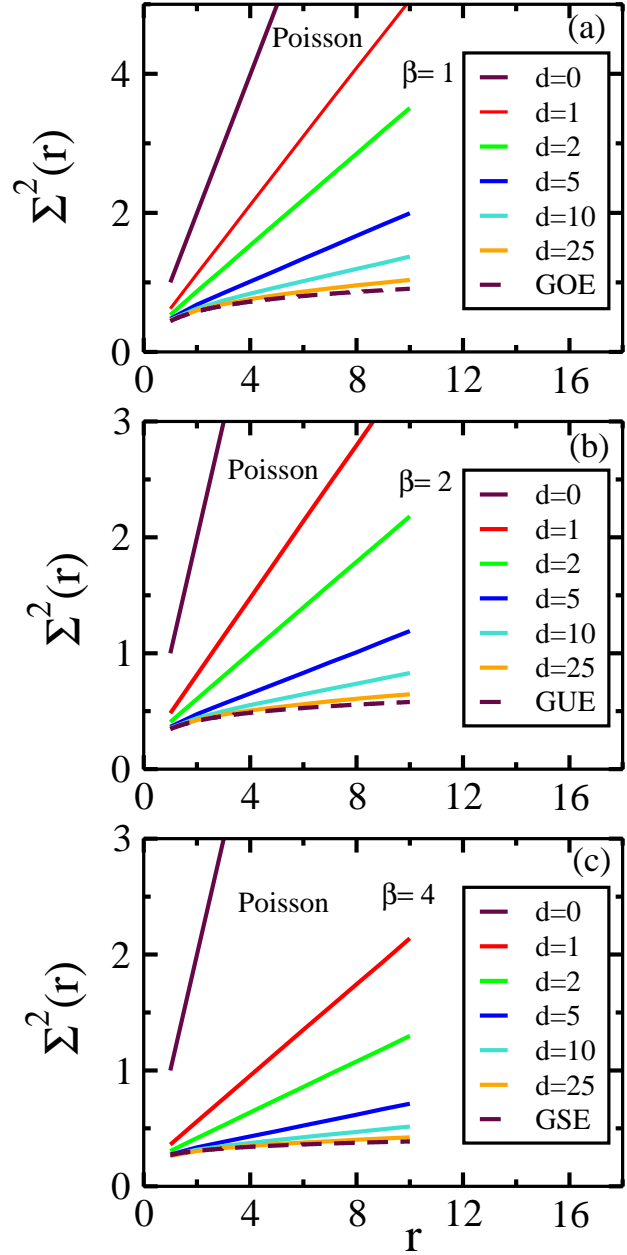}
\caption{Number variance [$\Sigma^2(r)$ vs. $r$] for FRCG models with $d=0,1,2,5,10,25$. We show results for (a) $\beta=1$, (b) $\beta = 2$, and (c) $\beta = 4$. The dashed line denotes the classical Gaussian ensemble result.}
\label{f7}
\end{figure}

Next we present some results for fluctuation measures for different $d$-values. Fig.~\ref{f4} corresponds to the nearest-neighbor spacing distribution for $d= 0,1,2,5,10$ and $\beta=1,2,4$. For comparison, the respective classical results have also been plotted. We can see that the MC data for $d=10$ is already very close to the classical result. In Fig.~\ref{f5}, we plot the two-point correlation functions [$R_2(s)$ vs. $s$] for $d=0,1,2,3,5$ for all three $\beta$-values. Fig.~\ref{f6} shows the two-point cluster function, which is defined as $Y_{2}(s)= 1-R_{2}(s)$. In Fig.~\ref{f7}, we show the number variance [$\Sigma^2(r)$ vs. $r$] for $d=0,1,2,5,10,25$ and $\beta=1,2,4$. For comparison, we have also plotted the classical random matrix theory (RMT) results. The transition from Poisson to Gaussian ensembles as $d$ is increased from 0 to $N-1$ is very clear.

\begin{figure}[H]
\centering
\includegraphics[width=0.6\textwidth]{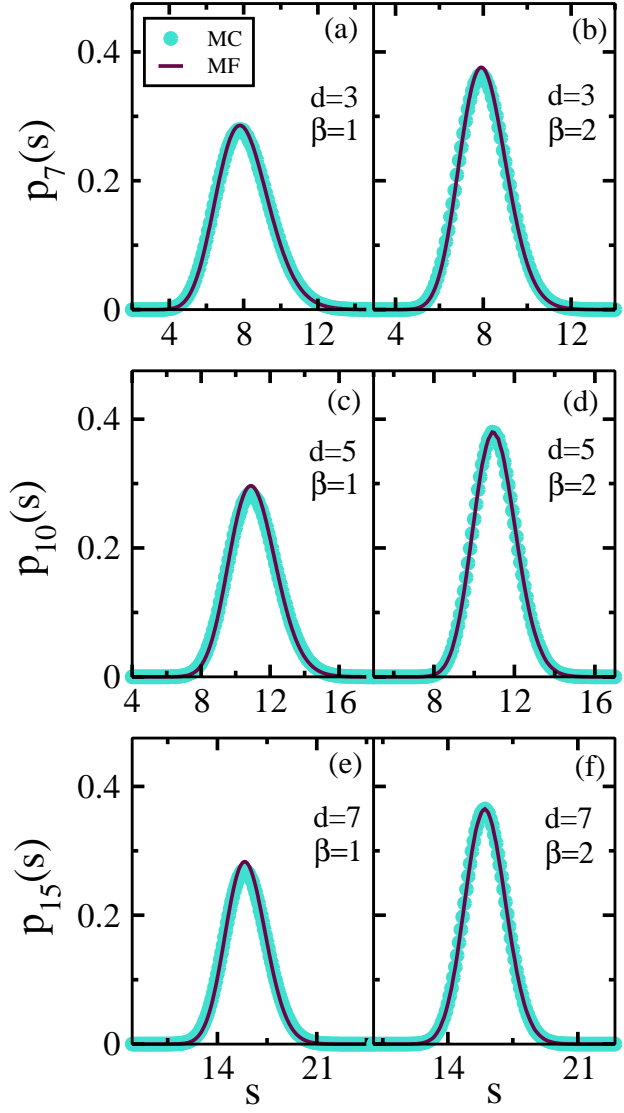}
\caption{Spacing density [$p_{n-1}(s)$ vs. $s$] for FRCG models. We consider various values of $d,\beta,n$ -- as indicated. The filled circles correspond to MC results, and the solid lines denote the MF result from Eq.~(91) of paper I.}
\label{f8}
\end{figure}

Before concluding this section, it is useful to compare our MC results with the mean-field (MF) results discussed in Sec.~9 of paper I. In Fig.~\ref{f8}, we plot the spacing density [$p_{n-1}(s)$ vs. $s$] for various values of $n,d$ and $\beta$. The MF result for $p_{n-1}(s)$ is given in  Eq.~(91) of paper I. In all cases shown, we find very good agreement between MF results and the exact MC results.

\section{Physical Applications of FRCG Models}
\label{s4}

Let us now demonstrate the applicability of the FRCG models to two important physical applications, i.e., the QKR and BRM. The QKR is a prototypical example of quantum chaotic systems. The spectral fluctuations of QKR in the strongly chaotic regime correspond to the classical random matrix or Wigner-Dyson statistics \cite{fi86,fi90}. This statistics can be obtained from infinite-range Coulomb gas models, as discussed in Sec.~2 of paper I \cite{fd62}. In the following sections, we demonstrate that spectral fluctuations of QKR are described by FRCG models. Earlier work has also shown a deep connection between QKR and BRM \cite{cci79,pf89}. Therefore, the statistics of BRM ensembles is also described by FRCG models. In this section, we introduce QKR and BRM and give their definitions.

\subsection{Quantum Kicked Rotors (QKR)}
\label{s4a}

Following Izrailev \cite{fi86}, we consider a finite-dimensional $\left[ N\times N\right]$ matrix model for QKR. The evolution operator is given by $U=BG$, where
\begin{equation} \label{e9}
B(\alpha)= \exp\left[-i \frac{\alpha}{\hbar} \cos(\theta +\theta_{0})\right],
\end{equation}
and
\begin{equation}\label{e10}
G=\exp\left[-\frac{i}{2\hbar}(p+\gamma)^{2}\right],
\end{equation}
with $\theta$ and $p$ being the position and momentum operators. Here, $\alpha$ is the kicking parameter, $\theta_{0}$ is
the parity-breaking parameter, and $\gamma$ is the time-reversal-breaking parameter $(0\leq\gamma<1)$. In position representation,
\begin{equation}\label{e11}
B_{mn}=\exp\left[-i\frac{\alpha}{\hbar}\cos\left(\frac{2\pi m}{N}+\theta_{0}\right)\right]\delta_{mn} ,
\end{equation}
\begin{equation}\label{e12}
G_{mn}=\frac{1}{N}\sum_{l=-N^{\prime}}^{N^{\prime}}\exp\left[-i\left(\frac{\hbar}{2}l^{2}-\gamma l-\frac{2 \pi \mu l}{N}\right)\right] ,
\end{equation}
where $N^\prime = (N-1)/2$. The indices $m,n=-N^{\prime},-N^{\prime}+1,\cdots,N^{\prime}$. Then, the evolution operator becomes
\begin{equation}\label{e13}
U_{mn} = \frac{1}{N}\exp\left[-i{\alpha}\cos\left(\frac{2\pi m}{N}+\theta_{0}\right)\right] \times \sum_{l=-N^{\prime}}^{N^{\prime}}\exp\left[-i\left(\frac{l^{2}}{2}-\gamma l-\frac{2 \pi \mu l}{N}\right)\right],
\end{equation}
where $\mu= m-n$. We have set $\hbar=1$. One knows that, when parity is broken ($\theta_{0}\neq 0$), and $\alpha^{2}\gg N \gg 1$, then the eigenvalue spectra of $U$ accurately exhibits classical random-matrix spectral fluctuations (e.g., spacing distribution, number variance). For $\gamma=0$, the fluctuations are characterized by $\beta = 1$ (GOE). For $\gamma \neq 0~(\gamma\gg N^{-3/2})$, the fluctuations obey $\beta=2$ (GUE) statistics \cite{fi86,prs73}.

The numerical results presented in this paper for the spectral statistics of QKR were obtained by studies of the matrix $U_{mn}$ in Eq.~(\ref{e13}). For the weakly-chaotic regime (small $\alpha$), we consider a single matrix of large size. For the strongly-chaotic regime (large $\alpha$), we study an ensemble of matrices generated for values of $\alpha$ in a small window around an average $\bar{\alpha}$. In the latter case, we will label the results by the value of $\bar{\alpha}$.

\subsection{Banded Random Matrices (BRM)}
\label{s4b}

Next, we introduce BRM ensembles $\{A\}$ of dimensionality $N$. The matrix $A$ is banded if $A_{jk}=0$ for $| j-k|  > b$, where $b$ is the bandwidth. The jpd of the matrix distribution for Gaussian BRM ensembles is
\begin{equation}\label{e15}
P(A) = C \exp(-\mbox{Tr} A^{2}/4v^{2}),
\end{equation}
with $v^{2}$ being the variance of the nonzero off-diagonal matrix elements. The matrices $A$ can be real symmetric, complex hermitian, or quaternion self-dual corresponding to $\beta=1,2,4$ respectively. For $b=N-1$, the GOE, GUE, GSE are recovered respectively. We can also generalize Eq.~(\ref{e15}) to the non-Gaussian case, where $A^{2}$ in the exponent is replaced by a positive-definite function of $A$. We will not discuss the non-Gaussian case here.

It can be shown that the eigenvalue density of BRM is semicircular, as in the classical ensembles \cite{wfl91}. The density is given by
\begin{equation}\label{e16}
\rho(x)= \frac{2\sqrt{R^{2}-x^{2}}}{\pi R^{2}} ,
\end{equation}
where the radius $R^{2}=8\beta b v^{2}$. However, the number of eigenvalues outside the semicircle increases as $b$ decreases. For example, for $N=1001$ the number of eigenvalues outside the semicircle is 1-2 for the classical case, as compared to 10-20 for the BRM with $b \simeq 50$.

There have been several important studies of the level statistics of BRM. Mirlin and Fyodorov (MF) \cite{mf91,fm91,fm93} have studied BRM analytically using the supersymmetric nonlinear sigma model. They demonstrated that this model exhibits localization on the scale $\ell \sim b^2$. They also showed that the nonlinear sigma model was equivalent to a 1-dimensional disordered wire with diffusion constant $D \sim \ell$. Numerical experiments by Casati et al. \cite{cmi90,cim91} confirmed the localization of eigenvectors of BRM on the scale $\ell$. Casati et al. \cite{cgi90} also showed a similar localization of eigenvectors in random matrices of quantum chaotic systems.

MF also proposed that the BRM ensemble with bandwidth ranging from $1 \rightarrow N-1$ is suitable for interpolating between the integrable regime (with Poisson statistics) and the chaotic regime (with classical RMT statistics) of time-reversal invariant quantum systems. We discuss these limits in the context of the diffusion constant $D$. The dynamics is diffusive in the limit $D/N = b^2/N \gg 1$. In this case, Altshuler and Shklovskii \cite{as86} have showed that the spectral statistics obey classical RMT. In the opposite limit $b^2/N \ll 1$, the dynamics is localized and we expect Poisson statistics to apply.

\subsection{Connection between QKR, BRM and FRCG Models}
\label{s4c}

It has empirically been shown by several authors \cite{fi90,cmi90,cim91,cgi90} that BRM and QKR give the same nearest-neighbor spacing density $p_{0}(s)$ when $b^{2}/N = \alpha^{2}/N$. These authors also showed that, in the momentum representation, the QKR matrix $U$ is banded.

Let us examine the structure of operators in the QKR evolution. We will first demonstrate that, even though $p$ and $\cos\theta$ are non-random operators, they have matrix properties analogous to Gaussian random matrix ensembles in $U$-diagonal representation. The randomness arises from the statistical properties of eigenvectors of $U(\alpha)$, which are similar to those of Gaussian ensembles.

\begin{figure}[H]
\centering
\includegraphics[width=0.6\textwidth]{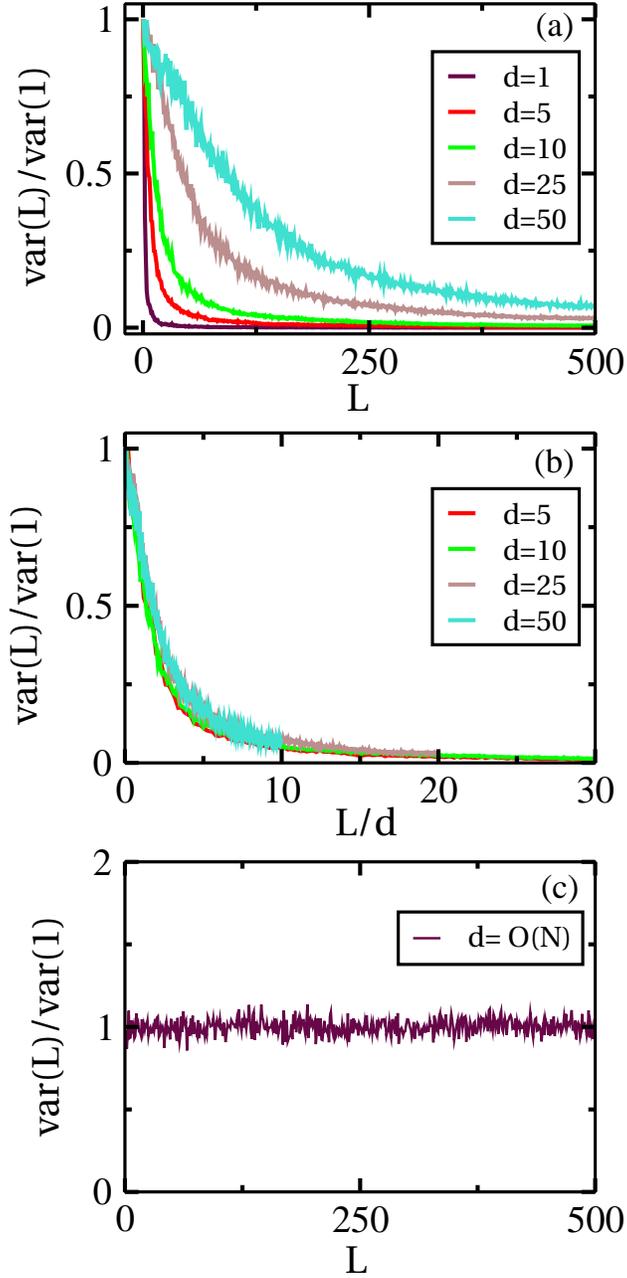}
\caption{(a) Plot of $\mbox{var}(L)/\mbox{var}(1)$ vs. $L$ for the QKR momentum operator in $U$-diagonal basis. We have $\gamma = 0$, and the values of $\alpha$ are such that $d = \alpha^2/N = 1,5,10,25,50$. (b) Data in (a), plotted against the scaled variable $L/d$. (c) Analogous to (a), but for large value of $\alpha$ so that $d = O(N)$.}
\label{f9}
\end{figure}

We first consider the case with $\gamma = 0$. The $\cos\theta$ operator in $U$-diagonal representation has matrix properties analogous to the GOE (not shown here). However, since its eigenvalues are fixed, there are weak correlations among different matrix elements. Similarly, when we write $p$ in $U$-diagonal representation, it has matrix properties analogous to a Gaussian BRM ensemble with width depending on $\alpha$. The matrix elements $p_{jk}$ in the position basis of $U$ are given by
\begin{align}
\label{pjk}
p_{jk} &\equiv  \langle \phi_{j}| p | \phi_{k} \rangle \nonumber \\
&=\sum_{m,n} \langle \phi_{j}| m \rangle \langle m | p | n \rangle \langle n | \phi_{k} \rangle .
\end{align}
Here, $| \phi_j \rangle$ represents the eigenfunctions of the evolution operator $U$, and $| m \rangle $ represents the basis of the momentum operator. Note that $p$ is a diagonal matrix in the self-basis. Eq.~(\ref{pjk}) can be simplified to
\begin{equation}
p_{jk}= \frac{1}{N}\sum_{l,m,n} l \hbar~e^{i2\pi l(m-n)/N} \langle \phi_{j}| m \rangle \langle n |  \phi_{k} \rangle ,
\end{equation}
where the eigenfunctions are ordered by the corresponding eigenvalues. At this stage, it is useful to introduce the parameter $d=\alpha^2/N$, which will shortly be identified as the range of the corresponding FRCG model. Let us first consider some values of $\alpha$ for which $d \ll N$. In Fig.~\ref{f9}(a), we plot the normalized variance of the off-diagonal elements of $p$, $\mbox{var}(L)/\mbox{var}(1)$ vs. $L$, where $L$ is the distance from the diagonal. We see that the variance decays rapidly with $L$, demonstrating that $p$ is banded. A very interesting property of $\mbox{var}(L)$ is that the decay rate scales linearly with $d$. This is shown in Fig.~\ref{f9}(b), where $\mbox{var}(L)/\mbox{var}(1)$ is plotted against $L/d$, resulting in a neat data collapse. Fig.~\ref{f9}(c) shows the corresponding plot for $\alpha = 1000$ so that $d = O(N)$. In this case, the $p$-matrix is no longer banded. For large $d$, the variance is $\mbox{var}(L) \simeq N/12$ \cite{prs73}.

For $\gamma = 0.7$ (which satisfies $\gamma \gg N^{-3/2}$ \cite{prs73}), the above scenario applies again for the operators $\cos\theta$ and $p$, with GOE-like matrices replaced by GUE-like matrices. The $p$-matrices are banded (on a scale $d=\alpha^2/N$) or extended, depending on the value of $\alpha$.

We have demonstrated above that BRM arise naturally in the study of QKR \cite{cmi90,cim91,cgi90}. Let us next examine some properties of BRM. We start with the observation that the sum of two BRM $B_{1}$ and $B_{2}$, with the same bandwidths $b$ and variances  $v_{1}^{2}$, $v_{2}^{2}$, is also a BRM of bandwidth $b$ with variance $v_{1}^{2}+ v_{2}^{2}$. We investigate the statistics of the matrix elements of $B_2$ in the $B_{1}$-diagonal representation. For simplicity, we consider the case with $\beta=1$. The cases $\beta=2, 4$ yield similar results.

\begin{figure}[H]
\centering
\includegraphics[width=0.55\textwidth]{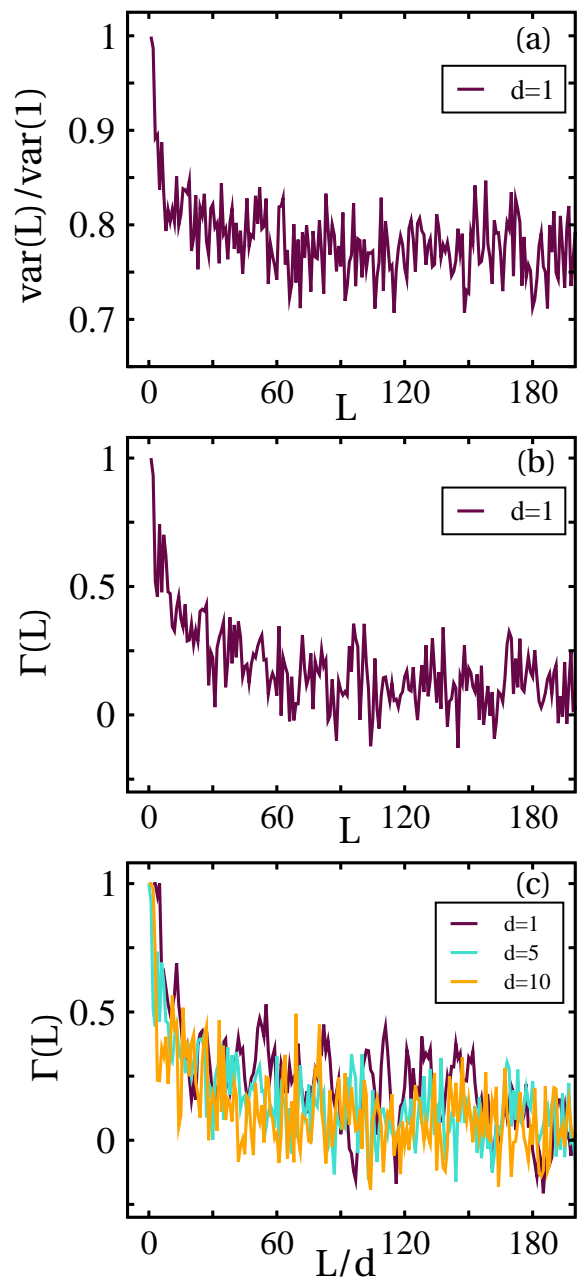}
\caption{(a) Plot of $\mbox{var}(L)/\mbox{var}(1)$ vs. $L$ for BRM. The bandwidth $b = 32$, so that $d=b^2/N=1$. (b) Plot of $\Gamma (L)$ vs. $L$ for the data in (a). (c) Superposition of $\Gamma (L)$ vs. $L/d$ for $b=32,72,100$ so that $d=1,5,10$.}
\label{f10}
\end{figure}

In Fig.~\ref{f10}(a), we plot the variance of $B_{2,jk}$ as a function of $L=|j-k|$. The bandwidth $b = 32$, so that $d=b^2/N=1$. We observe that the variance decays rapidly as in Fig.~\ref{f9}(a), but settles to a non-zero value $K$. This constant is approximately $\mbox{Tr} B_{2}^{2}/N^{2}$ for large $N$, and arises due to the semi-circular level density. In Fig.~\ref{f10}(b), we plot
\beq
\label{gammal}
\Gamma (L) = \frac{\mbox{var}(L)-K}{\mbox{var}(1)-K}
\eeq
as a function of $L$ for the data in Fig.~\ref{f10}(a). The decay rate is again proportional to $d$, which is confirmed by plotting $\Gamma (L)$ vs. $L/d$ for $d=1,5,10$ in Fig.~\ref{f10}(c).

\begin{figure}[H]
\centering
\includegraphics[width=0.6\textwidth]{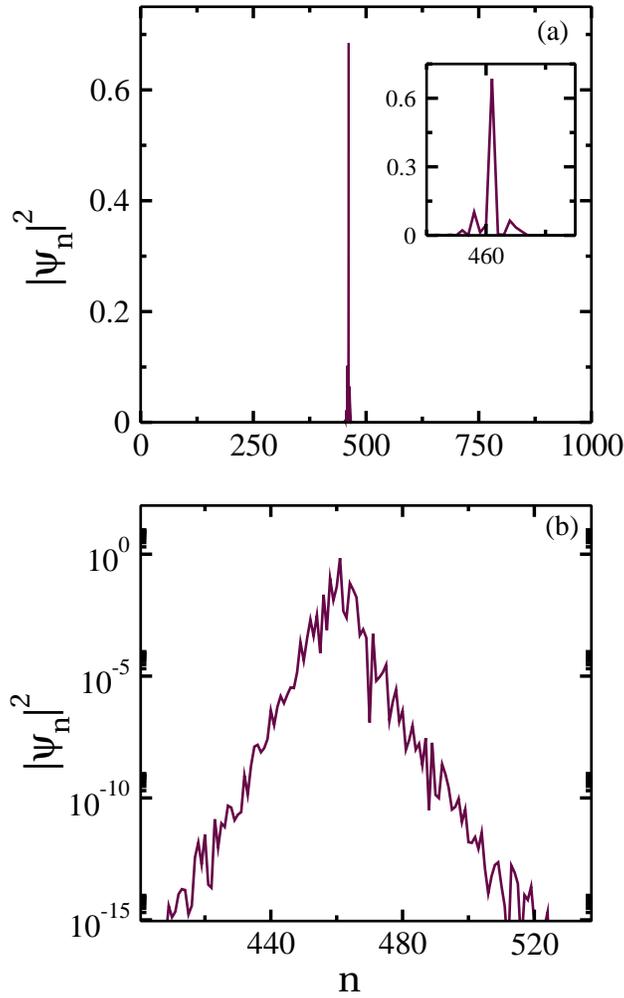}
\caption{Typical eigenvector of a BRM with bandwidth $b=5$. (a) Plot of $|\psi_n|^2$ vs. $n$ on a direct scale. Here, $n$ is the component of the eigenvector. The inset shows an expanded region near the point of localization. (b) Data in (a), plotted on a linear-log scale. The eigenfunction decays exponentially from the point of localization.}
\label{f11}
\end{figure}

\begin{figure}[H]
\centering
\includegraphics[width=0.6\textwidth]{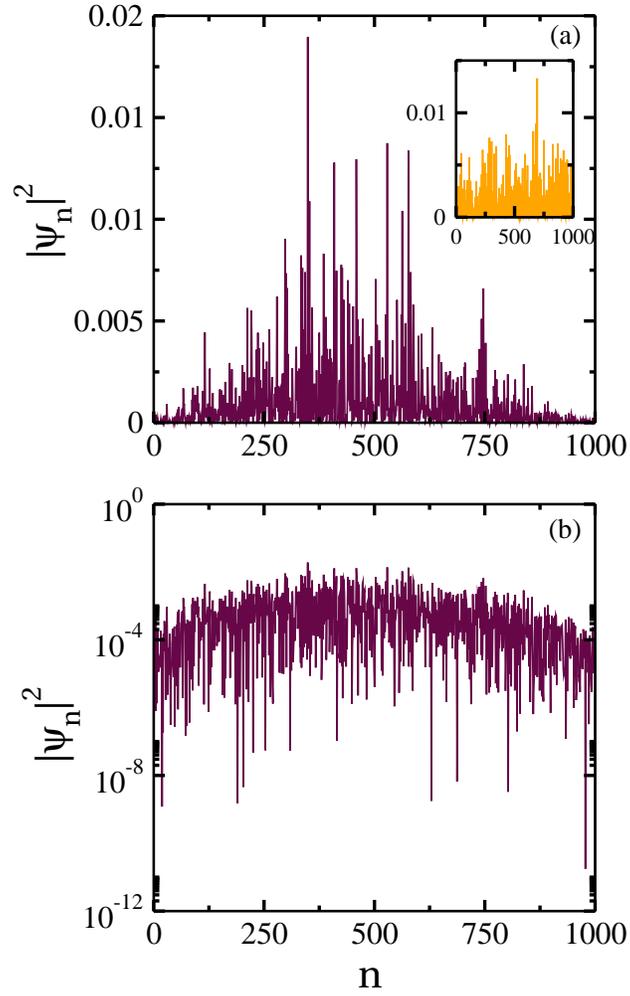}
\caption{Typical eigenvector of a BRM with bandwidth $b=250$ so that $d=63$. (a) Plot of $|\psi_n|^2$ vs. $n$. This corresponds to an extended state. The inset shows the eigenvector when $d=N-1$. (b) Data in (a), plotted on a linear-log scale.}
\label{f12}
\end{figure}

Next we turn our attention to the structure of eigenvectors of BRM. In Fig.~\ref{f11}(a), we plot $| \psi_{n}| ^{2}$ vs. $n$, where $\psi_{n}$ is the component of a typical eigenvector of $B_{2}$ with $b=5$. It is sharply localized around a particular value of $n$. In Fig.~\ref{f11}(b), we plot $| \psi_{n}| ^{2}$ vs $n$ on a linear-log scale. This plot shows that the decay of eigenvectors is exponential in the distance from the peak. This localization disappears as $d$ increases. In Fig.~\ref{f12}, we plot a typical $| \psi_{n}| ^{2}$ vs. $n$ for $b=250$, and see that the eigenvector is extended.

Where do FRCG models fit into the above framework? We have proposed recently \cite{pkp17} that QKR and BRM can be modeled by FRCG with range
\begin{equation}\label{e14}
d = \alpha^{2}/N = b^{2}/N ,
\end{equation}
valid for all fluctuation measures. As we will see shortly, both the diffusive and localized limits discussed in Sec.~\ref{s4b} are realized in our FRCG models. We should emphasize that the range $d$, as defined in Eq.~(\ref{e14}), can also take non-integer values. Therefore, we will subsequently introduce FRCG models with fractional values of $d$. This will facilitate a better understanding of the Poisson $\rightarrow$ Wigner-Dyson crossover as $d$ goes from 0 to $N-1$.

\section{Spectral Fluctuations for QKR and BRM and Comparison with FRCG Models}
\label{s5}

In this section, we present a detailed comparison between our FRCG results (obtained by analytic/MC approaches) and numerical results for QKR/BRM. In our subsequent discussion, the term ``theory'' will refer to the FRCG results. For the QKR, we consider the matrices $U_{nm}$. For the BRM, we study banded matrices with elements of the GOE-type or GUE-type. It is useful to summarize here various parameters for our numerical studies. The numerical results for QKR shown below correspond to $N = 1001$ (unless otherwise stated), $\theta_{0} = \pi/2N$, and $\gamma = 0.0, \gamma = 0.7$. We will show results for several values of the kicking parameter $\alpha$. Our numerical results for BRM correspond to $N=1001$, and several values of the bandwidth $b$.

Before presenting results for spectral fluctuations, it is important to clarify a technical detail about the procedure we adopt for unfolding of the eigenvalue spectrum. For the QKR, the level density of eigenvalues is uniform. Therefore, a multiplication factor of $N/2\pi$ makes the average density unity everywhere. As mentioned in Sec.~\ref{s4}, BRM have the additional feature that the level density is semicircular. In this case, for unfolding purposes, we use the radius given after Eq.~(\ref{e16}).

\begin{figure}[H]
\centering
\includegraphics[width=0.6\textwidth]{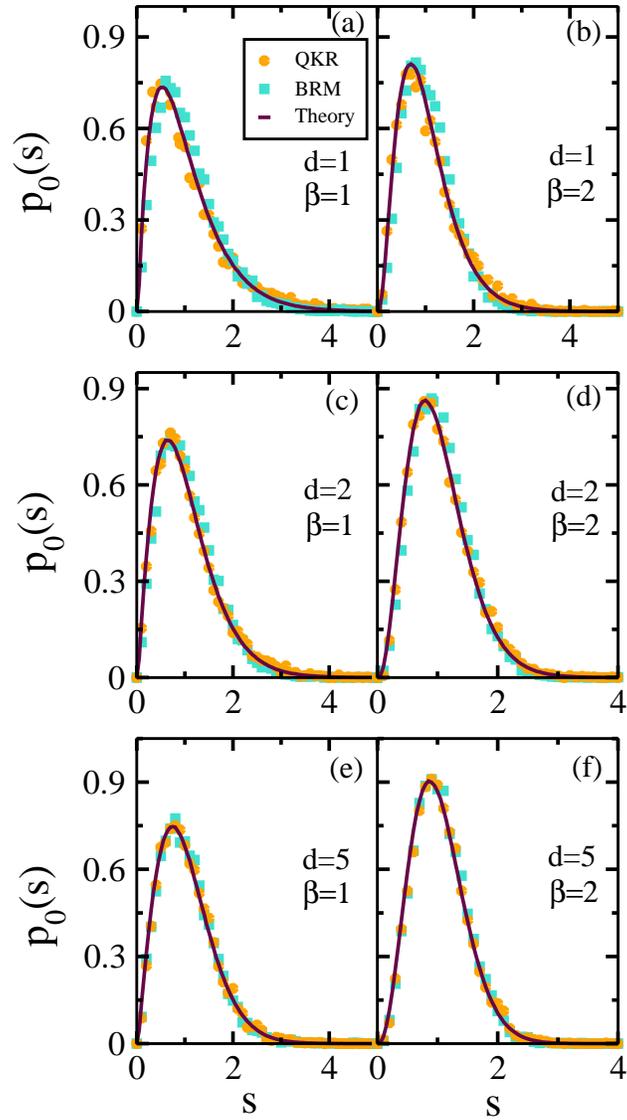}
\caption{Spacing density $p_{0}(s)$ vs. $s$ for QKR and BRM. The solid line denotes the FRCG result. We show data for various values of $d$ and $\beta$, as indicated.}
\label{f13}
\end{figure}

\begin{figure}[H]
\centering
\includegraphics[width=0.6\textwidth]{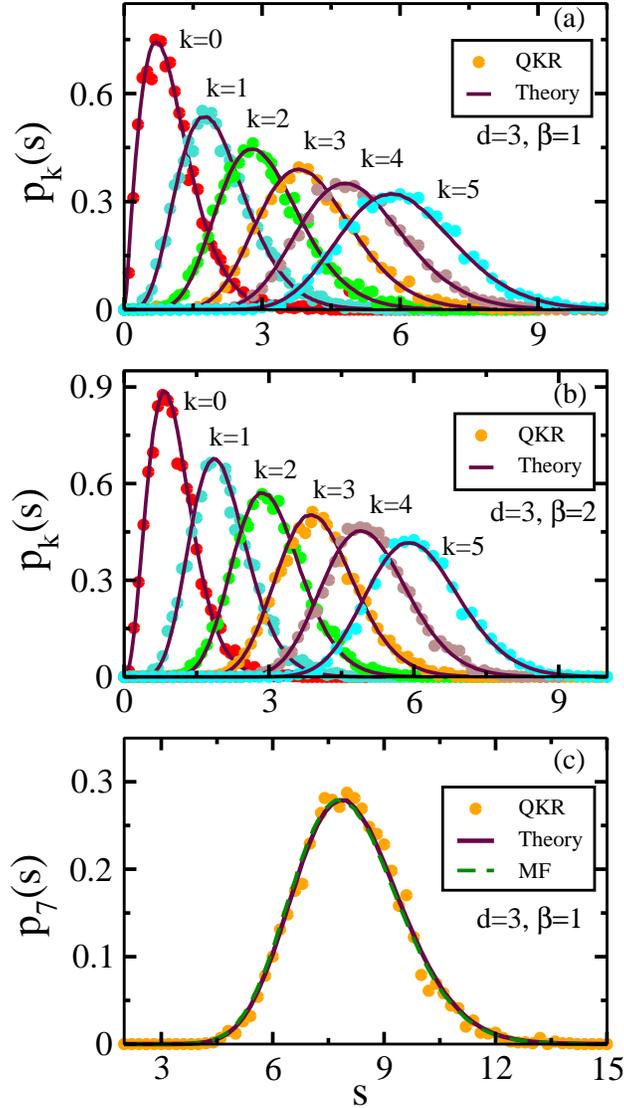}
\caption{Spacing density $p_{n-1=k}(s)$ vs. $s$ for QKR. The solid line denotes the FRCG result. (a) Data for $d=3,\beta=1$ with $k=0,1,2,3,4,5$. (b) Data for $d=3,\beta=2$ with $k=0,1,2,3,4,5$. (c) Data for $d=3,\beta=1$ with $k=7$. We also superpose the MF result.}
\label{f14}
\end{figure}

In Fig.~\ref{f13}, we show results for $p_0(s)$ vs. $s$ for QKR and BRM. In Fig.~\ref{f14}, we plot the higher-order spacing distribution  $p_{k}(s)$ for $d=3$ and $\beta=1,2$ in QKR. In both figures, the agreement between QKR/BRM  and FRCG results is excellent. In Fig.~\ref{f14}(c), we have also plotted the MF result from Sec.~9 of paper I. Note that the results for $d=0,1,2$ are obtained analytically in paper I. The FRCG results for $d=3,5$ are obtained via the MC technique.

\begin{figure}[H]
\centering
\includegraphics[width=0.6\textwidth]{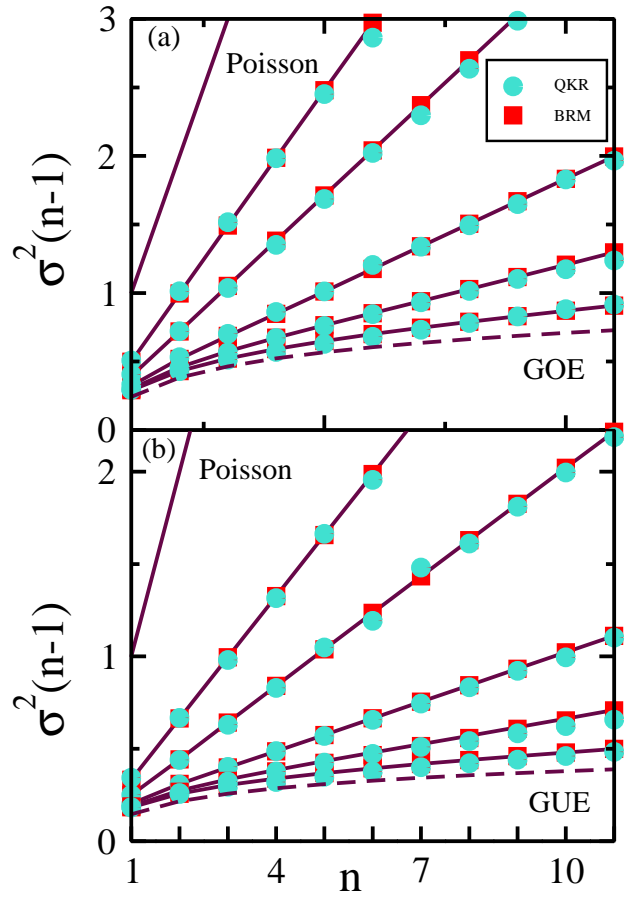}
\caption{Spacing variance $\sigma^2(n-1)$ vs. $n$ for QKR and BRM. We show data for $d=0,1,2,5,10,25,N-1$ (from top to bottom), and (a) $\beta=1$, (b) $\beta=2$. The solid lines denote the corresponding FRCG results. The dashed lines denote the classical ensemble results.}
\label{f15}
\end{figure}

In Fig.~\ref{f15}, we show the spacing variance for $\sigma^2 (n-1)$ vs. $n$ for QKR/BRM with $N=5001$. We have considered several values of $d$ in both GOE and GUE. As before, we compare our QKR/BRM results with theory from FRCG models, and see that the agreement is excellent. We make the following observations: \\
(a) $\sigma^{2}(n)$ vs. $n$ is linear for both GOE and GUE for $d = O(1)$. For $d=O(N)$, $\sigma^{2}(n)$ shows logarithmic behavior as in classical statistics. \\
(b) The spacing variance in the $\beta=2$ case is roughly half of that in the $\beta=1$ case for each value of $n$. \\
These behaviors are qualitatively similar to our analytical results for FRCG models in paper I.

\begin{figure}[H]
\centering
\includegraphics[width=0.6\textwidth]{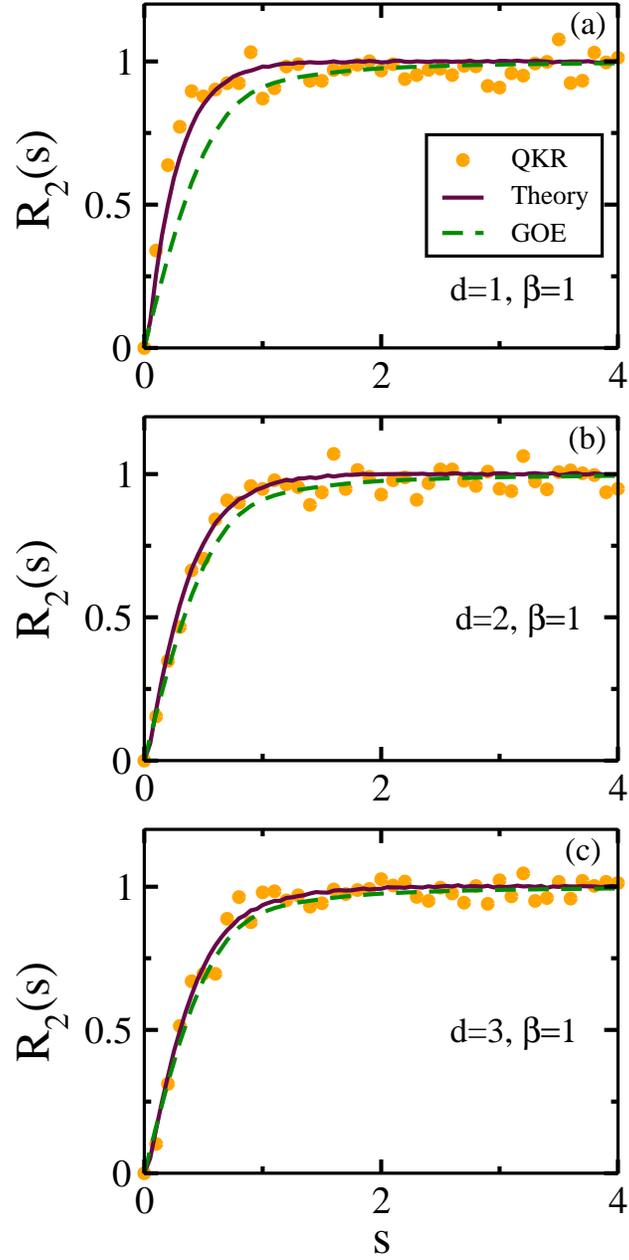}
\caption{Two-point correlation function $R_{2}(s)$ vs. $s$ for QKR. The solid lines denote the corresponding FRCG results. The dashed lines denote the GOE result. We show data for $\beta=1$ and (a) $d=1$, (b) $d=2$, (c) $d=3$.}
\label{f16}
\end{figure}

\begin{figure}[H]
\centering
\includegraphics[width=0.6\textwidth]{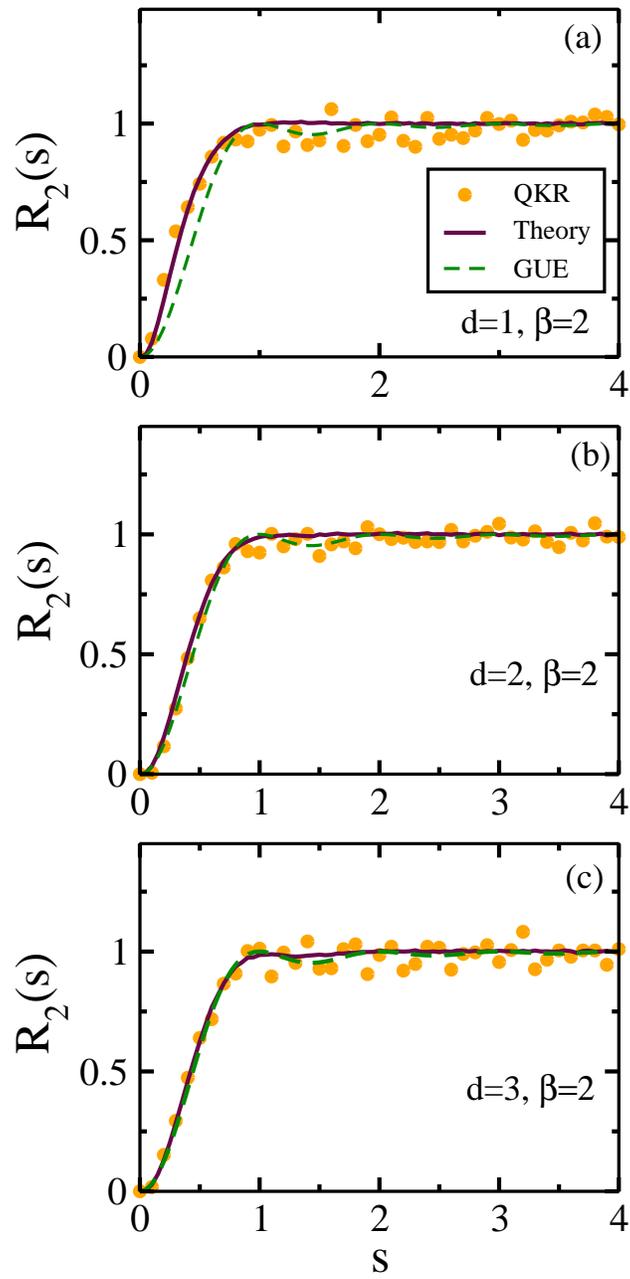}
\caption{Analogous to Fig.~\ref{f16}, but for the $\beta=2$ case.}
\label{f17}
\end{figure}

In Figs.~\ref{f16} and \ref{f17}, we plot $R_2(s)$ vs. $s$ for QKR with $\beta=1,2$, and compare with FRCG results. Again, there is very good agreement between QKR and theory. Notice that our results are already very close to the relevant classical result (GOE or GUE) for $d=3$. Thus, there is a rapid transition from Poisson ($d=0$) to classical ($d=N-1$) results.

A more quantitative comparison with physical systems may require the introduction of Coulomb gas models where the interaction strength decays gradually with distance rather than the sharp cut-off considered here. This will be part of our future investigation of this problem.

\section{Fractional Values of $d$: Comparison with QKR}
\label{s6}

Our discussion so far has focused on integer values of $d$. However, the parameter $d$ relevant for QKR $(d=\alpha^{2}/N)$ and BRM $(d=b^{2}/N)$ can take non-integer values also. For a complete description of QKR and BRM with arbitrary $d$, we introduce an FRCG model with fractional $d$. We generalize the jpd for nearest-neighbor spacings in Sec.~7 of paper I as follows:
\begin{equation}\label{e17}
P_{d}(s_1,\cdots,s_N) = C_d \delta\left(\sum_{i=1}^{N}s_{i}-N\right) \prod_{j=1}^{N}\prod_{k=0}^{[d]}(s_{j}+\cdots+s_{j+k})^{\beta\Delta(k)},
\end{equation}
where $[d]$ is the largest integer $\leq d$. Moreover, $\Delta(k)=1$ for $k=0,1,\cdots, [d]-1$, and $\Delta([d])=d-[d]$. We make the following observations: \\
(i) For integer $d$, Eq.~(\ref{e17}) reduces to the definition of $P_{d}$ in Eq.~(56) of paper I. \\
(ii) All cases with $0 \leq d \leq 1$ are analytically tractable as there are only one-particle terms in $P_d$. This case is analogous to the $d=0,1$ cases in Sec.~8 of paper I. The corresponding nearest-neighbor distribution is 
\begin{equation}\label{e18}
p_{0}(s)= \frac{(\beta d +1)^{(\beta d +1)}}{\Gamma(\beta d+1)}s^{\beta d} e^{-(\beta d+1)s}.
\end{equation}
The $(n-1)^{\rm th}$ spacing distribution is given by
\begin{equation}\label{e19}
p_{n-1}(s)=\frac{(\beta d+1)^{(\beta d+1)n}}{\Gamma((\beta d+1)n)} s^{(\beta d+1)n-1}e^{-(\beta d+1)s}.
\end{equation}
(iii) Our choice of $\Delta (k)$ ensures that the MF approximation of $P_{d}$ yields the results in Eqs.~(\ref{e18})-(\ref{e19}) even for non-integer $d$.

\begin{figure}[H]
\centering
\includegraphics[width=0.6\textwidth]{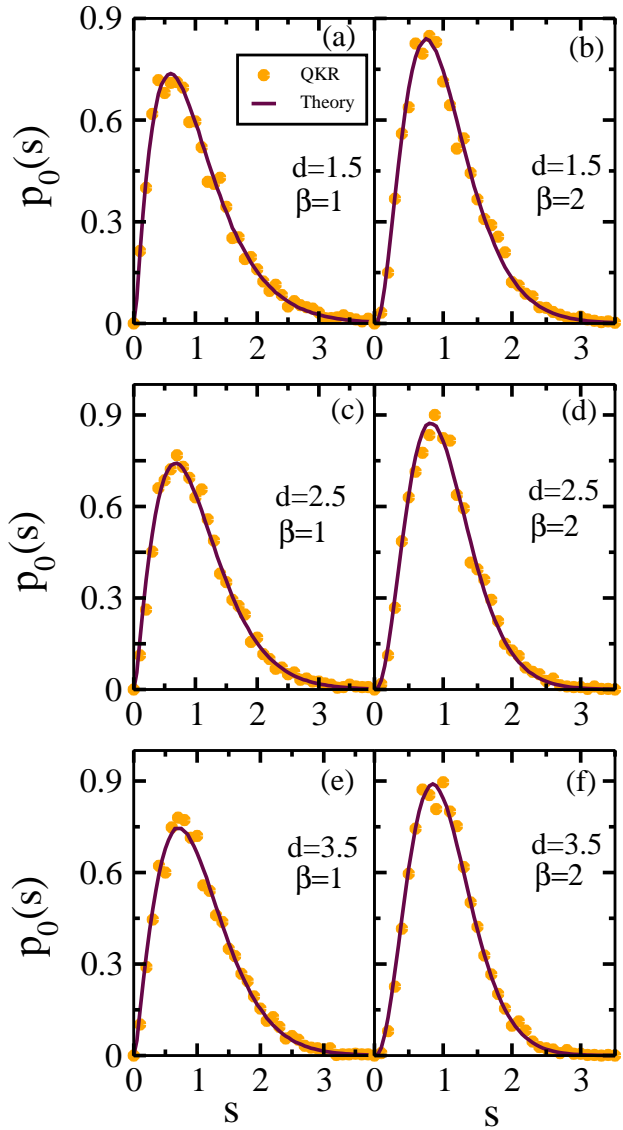}
\caption{Plot of $p_0(s)$ vs. $s$ for QKR with fractional values of $d$. The solid lines denote the appropriate FRCG result. We show results for various values of $\beta$ and $d$, as indicated.}
\label{f18}
\end{figure}

Next, we present some results for non-integer $d$ in Fig.~\ref{f18}. The theoretical results are obtained via MC in this case. We plot $p_{0}(s)$ vs. $s$ for QKR and FRCG models with $d=1.5,2.5,3.5$ and $\beta=1,2$. The excellent agreement confirms the applicability of our fractional FRCG model to understand the spectral statistics of QKR.

\section{Summary and Discussion}
\label{s7}

Let us conclude this paper with a summary and discussion of our results in this two-part exposition on {\it finite-range Coulomb gas} (FRCG) models and their application in physical systems.

In paper I, we introduced FRCG models as a natural generalization of Dyson's Brownian motion models for eigenvalue spectra of random matrix ensembles. These are parametrized by the range of interactions between eigenvalues, denoted as $d$. Our FRCG models provide a novel route for transition from Poisson statistics (for $d=0$) and classical random matrix statistics (for $d=N-1$, where $N$ is the dimensionality of the matrices). In paper I, we also introduced an integral-equation approach for analytical solution of these FRCG models. The integral equation is analytically tractable for $d \le 2$. However, for $d>2$, the equations become increasingly complicated. For $d>2$, we have proposed a mean-field (MF) approximation, which yields simple and accurate solutions. In this paper (II in the series), we have also proposed a Monte Carlo (MC) technique which yields precise results for spectral statistics of FRCG models. The MC technique is validated by comparison with analytical results, wherever these are available. We use the term ``theory'' to describe exact analytic and MC results for FRCG models.

It is natural to ask whether the elegant framework of FRCG models has useful physical applications. This is the primary focus of the present paper, where we have demonstrated that the eigenvalue statistics of {\it quantum kicked rotors} (QKR) and {\it banded random matrices} (BRM) is described by FRCG models. The QKR are characterized by a kicking parameter $\alpha$, which describes how chaotic the system is. The BRM are parametrized by the bandwidth $b$, which is the off-diagonal distance upto which the matrix has non-zero entries. Earlier work has shown that QKR and BRM yield the same results for $p_0(s)$ if $\alpha^2/N = b^2/N$. In this paper II, we have shown that FRCG models with $d = \alpha^2/N = b^2/N$ provide a framework for deriving the spectral properties of QKR/BRM. We have presented results from a detailed comparison of diverse spectral properties in QKR/BRM and FRCG. In all cases, the agreement is excellent.

The QKR has been a fundamental paradigm in the area of quantum chaos. Therefore, it is gratifying to see that FRCG models provide an excellent description of the QKR statistics. An important direction for future research is the identification of other physical systems which are modeled by FRCG. There are many systems which exhibit a crossover from Poisson to classical statistics as a parameter is varied. In this context, there have been studies of diverse systems such as atomic spectra \cite{rp60}, random matrix models \cite{ap81},  quantum chaotic systems \cite{wm88,svz84}, Anderson localization \cite{sm16,bks11,mmm97}, quark-gluon plasma \cite{kp10} and neural networks \cite{ahn16}. Clearly, there are several different routes whereby this transition can be realized. It is our belief that the FRCG scenario may find application in several of these systems.

\newpage

\end{document}